\documentclass[iop]{emulateapj}
\usepackage{amsmath}
\usepackage{amssymb}
\usepackage{mathbbol}
\usepackage{dsfont}

\DeclareGraphicsExtensions{.eps,.eps.gz,.epsi}

\newcommand{\teff}{\mbox{$T_{\rm eff}$}} \newcommand{\logg}{{\rm{log}~$g$}}
\newcommand{\feh}{{\rm [Fe/H]}} 
\newcommand{\ebv}{$E(B-V)$}

\shorttitle{Stellar loci I. Metallicity dependence and intrinsic widths}
\shortauthors{Yuan et al.}
\begin{document}
%\begin{CJK*}{UTF8}{gbsn}

\title{Stellar loci I. Metallicity dependence and intrinsic widths}

\author
{Haibo Yuan\altaffilmark{1, 2},
Xiaowei Liu\altaffilmark{3, 1},
Maosheng Xiang\altaffilmark{3},
Yang Huang\altaffilmark{3},
Bingqiu Chen\altaffilmark{3}
}
\altaffiltext{1}{Kavli Institute for Astronomy and Astrophysics, Peking University, Beijing 100871, P. R. China; email: yuanhb4861@pku.edu.cn}
\altaffiltext{2}{LAMOST Fellow}
\altaffiltext{3}{Department of Astronomy, Peking University, Beijing 100871, P. R. China; email: x.liu@pku.edu.cn}
%\end{CJK*}

\journalinfo{submitted to The Astrophysical Journal}
\submitted{Received ; accepted }

\begin{abstract}
Stellar loci are widely used for selection of interesting outliers,
reddening determinations, and calibrations.
However, hitherto the dependence of stellar loci on metallicity has not been fully explored 
and their intrinsic widths are unclear.
In this paper, by combining the spectroscopic and re-calibrated imaging data of the SDSS Stripe 82, 
we have built a large, clean sample of dwarf stars with accurate colors and well determined metallicities 
to investigate the metallicity dependence and intrinsic widths of the SDSS stellar loci. 
Typically, one dex decrease in metallicity causes 0.20 and 0.02\,mag decrease in colors $u-g$ and $g-r$, 
and 0.02 and 0.02\,mag increase in colors $r-i$ and $i-z$, respectively.
The variations are larger for metal-rich stars than for metal-poor ones, and for F/G/K stars than for A/M ones.
Using the sample, we have performed two dimensional polynomial fitting to the $u-g$, $g-r$, $r-i$, and $i-z$ colors 
as a function of color $g-i$ and metallicity \feh.
The residuals, at the level of 0.029, 0.008, 0.008 and 0.011 mag for the $u-g$, $g-r$, $r-i$, and $i-z$ colors,
respectively can be fully accounted for by the photometric errors and metallicity uncertainties, suggesting that the intrinsic widths
of the loci are at maximum a few mmag. 
The residual distributions are asymmetric, revealing that a significant fraction of stars are binaries. 
In a companion paper, we will present an unbiased estimate of the binary fraction for field stars.
Other potential applications of the metallicity dependent stellar loci are briefly discussed.
\end{abstract}
\keywords{methods: data analysis -- stars: fundamental parameters -- stars: general -- surveys}

\section{Introduction} 
Stellar color locus (Covey et al. 2007; Davenport et al. 2014; Chen et al. 2014), 
defined by the distribution of main sequence stars, the majority of all stars, in the color-color space, 
is one of the most useful tools in astronomy.
It plays an essential role at least in two types of study:
a) the selection of candidates of various types of interesting object of unusual colors 
for followup spectroscopic studies,  such as 
quasars, white dwarfs, white-dwarf-main-sequence binaries, UV/IR excess objects,
blue horizontal branch stars, red clump stars, red giants, and young stellar objects;
and b) the determinations of the interstellar extinction and reddening using
the Rayleigh-Jeans color excess method (Majewski et al. 2011), the blue tip method (Schalfly et al. 2010) and
the spectral energy distribution (SED) fitting method (Berry et al. 2012; Green et al. 2014; Chen et al. 2014).
In addition, it has also been used for the photometric calibration of large scale imaging 
surveys (Ivezi{\'c} et al. 2007; High et al. 2009; Kelly et al. 2014).

Stellar colors depend on the stellar effective temperature, 
but also to a fair degree on metallicity remarkably, in particularly the blue colors. 
However, hitherto the lack of metallicity information and extremely accurate photometric colors for 
large samples of stars has prevented a quantitative study of the  
dependence of stellar loci on metallicity and their intrinsic widths, 
and, as a consequence, restricted the potential applications of the stellar loci as an extremely powerful astronomical tools.

The repeatedly scanned equatorial Stripe 82  ($|{\rm Dec}| < 1.266\degr$, 20$^h$34$^m$ $< {\rm Ra} <$ 4$^h$00$^m$) 
of the Sloan Digital Sky Survey (SDSS; York et al. 2000) has delivered accurate photometry 
for about one million stars in $u,g,r,i,z$ bands (Ivezi{\'c} et al. 2007).
This is the largest data set available with optical photometry internally consistent at the 1 per cent level, 
providing ``a practical definition of the SDSS photometric system" (Ivezi{\'c} et al. 2007).
In addition, over 40,000 stellar spectra have been obtained in this region. 
The basic stellar parameters, radial velocity, effective temperature, surface gravity, and metallicity, 
derived from the spectra and released with the SDSS Data Release 9 (DR9; Ahn et al. 2012), 
have been determined with the SEGUE Stellar Parameter Pipeline (SSPP; Lee et al. 2008a,b) developed for 
the {\rm Sloan Extension for Galactic Understanding and Exploration} ({\rm SEGUE}; Yanny et al. 2009).
Using an innovative spectroscopy based stellar color regression (SCR) method, 
Yuan et al. (2014a) have further 
re-calibrated the Stripe 82 data, achieving an unprecedented internal accuracy of 
about 0.005, 0.003, 0.002, and 0.002\,mag in the $u-g$, $g-r$, $r-i$, and $i-z$ colors, respectively. 
By combining the spectroscopic information and re-calibrated photometry of Stripe 82, 
we have constructed a large, clean sample of main sequence stars with 
well determined metallicities and extremely accurate colors. 
In this paper, the sample is used to investigate the metallicity dependence of 
stellar loci in the SDSS colors and their intrinsic widths.

The paper is organized as follows.  In Section 2, we introduce the spectroscopic and photometric data
used. The results are presented in Section\,3.
The conclusions are given in Section\,4, along with a brief discussion of some 
of the potential applications of  the newly deduced metallicity dependent stellar loci.

\section{Data}

We first select stars from the SDSS DR9 in Stripe 82 
that have been targeted spectroscopically with  
a spectral signal-to-noise ratio S/N $>$ 10 and stellar parameters determined with the SSPP (Lee et al. 2008a,b). 
A total number of 46,053 stars are selected.
We then cross-match this sample with the recalibrated photometric catalogs of Stripe 82 (Ivezi{\'c} et al. 2007; Yuan et al. 2014a) 
to obtain their photometric colors, with a matching radius of one arcsec. 
This yields 34,906 stars\footnote{The photometric catalogs of Stripe 82 compiled by Ivezi{\'c} et al. contain
only non-variable point sources of good photometry (See their Section 2.4), thus exclude 
a fraction of stars spectroscopically targeted by the SDSS.}.
The stars are then dereddened using the dust reddening map from 
Schlegel et al. (1998) and the empirical reddening coefficients of Yuan et al. (2014a), 
derived using a star pair technique (Yuan, Liu \& Xiang 2013).
Finally, stars of a line-of-sight extinction \ebv~$\le$ 0.15\,mag, \logg~$\ge$ 3.5\,dex, $0.3 \le (g-i)_0 \le 1.6$\,mag, 
\teff~$\ge$ 4,300\,K, and $-2.0 \le \feh~\le 0.0$\,dex are selected.
% dist_z le 600 pc 
The final sample contains 24,492 stars.
Their distribution in the $(g-i)_0$ and \feh~plane is shown in the top left panel of Fig.\,1. 
The photometric errors as a function of the observed magnitudes are also shown in Fig.\,1.
For $u$ band, the photon counting noises start to dominate the errors for 
$u \ga 19$. The error are about 0.01 mag at $u=19.0$\,mag and 0.1 mag at $u=22$\,mag.
For $g$, $r$, and $i$ bands, the errors are dominated by the calibration uncertainties and essentially 
constant, at a level of $0.006\pm0.001$, $0.005\pm0.001$, and $0.005\pm0.001$ mag, respectively.
For $z$ band, the photon counting noises dominate at $z \ga 18$\,mag, and the errors
are about 0.01 mag at $z=18.2$\,mag and 0.02 mag at $z=19$\,mag.

\begin{figure}
\includegraphics[width=90mm]{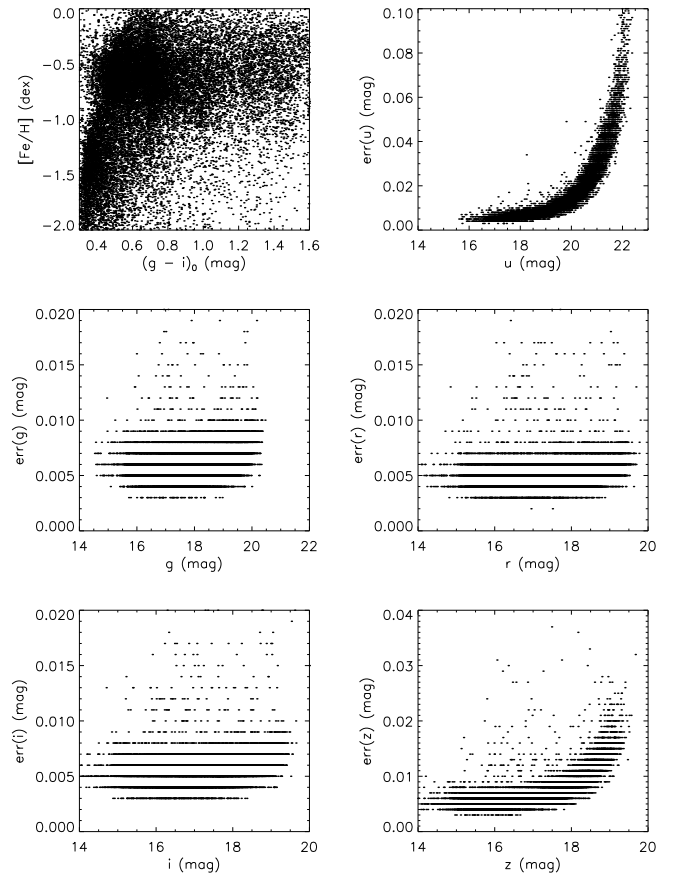}
\caption{
Distribution of [Fe/H] against dereddened color $(g-i)_0$
for the selected SDSS/DR9 stellar spectroscopic samples of Stripe 82 (top left panel)
and their photometric errors as a function of the observed magnitudes. 
}
\label{}
\end{figure}

\section{Results} 
Using the sample above, we have carried out a global two dimensional polynomial fit to $u-g$, $g-r$, $r-i$, and $i-z$ 
colors as a function of color $g-i$ and metallicity \feh. 
Note that all colors referred to hereafter are dereddened intrinsic values. 
For color $u-g$, a four-order polynomial with 15 free parameters is adopted.
For colors $g-r$, $r-i$, and $i-z$, a 3rd-order polynomial of 10 free parameters is used.
Two-sigma clipping is performed during the fitting process. 
The resultant fitting coefficients are listed in Table\,1.
Note the coefficients for $g-r$ and $r-i$ colors are fully degenerate.
The fitting residuals as a function of $g-i$ and \feh~are shown in Fig.\,2. 
The median values and standard deviations are marked in red. 
The scatter is bigger for redder stars. This is because redder stars are usually 
fainter and thus have larger photometric errors, especially in $u$ band (Fig.\,1). 

Fig.\,3 shows the differences between the fitted stellar loci of different metallicities ranging from \feh~=~$-2$ -- 0\,dex 
with respect to the corresponding ones of $\feh~= -0.5$\,dex. Lines of different colors indicate different metallicities.
As expected, the $u-g$ color is most sensitive to the metallicity. 
The $g-r$, $r-i$, and $i-z$ colors also show modest dependence on the metallicity. 
Typically, one dex decrease in metallicity leads to 0.20 and 0.02 mag decrease in colors $u-g$ and $g-r$,
and 0.02 and 0.02 mag increase in colors $r-i$ and $i-z$, respectively. 
The variations are larger for metal-rich and F/G/K stars than those of metal-poor stars and A/M stars. 
The trends are consistent with the synthetic colors predicted by the Kurucz model spectra (Lenz et al. 1998).
For comparison, the stellar loci of Covey et al. (2007) are also over-plotted in grey.
The loci of Covey et al. (2007) fall between those of $\feh~= -0.5$ and $\feh~= -1.0$, suggesting 
that most stars used by Covey et al. are from the thick disk, as expected.

To investigate the intrinsic widths of stellar loci, we divide the sample into 
bins of color and metallicity. For each bin, a Gaussian is used to 
fit the distribution of fit residuals to estimate the offset and dispersion of the residuals. 
To minimize the effects of photometric errors,
only stars of errors smaller than 0.01 mag in $u$ band are used in the case of color $u-g$
and likewise in $z$ band in the case of color $i-z$.  
The results for three typical color bins (centered at $g-i$ = 0.4, 0.9, and 1.4\,mag) and two metallicity 
bins (centered at $\feh = -0.3$ and $-1.0$\,dex) are plotted in Fig.\,4.  
The values of $g-i$ and [Fe/H] of the bin centers are labelled at the top of each panel. The bin widths are 0.4 mag 
and 0.4 dex in $g - i$  and [Fe/H], 
respectively. The Gaussian fits are over-plotted in red. 
The offset and dispersion of the Gaussian are labelled.

\begin{figure}
\includegraphics[width=90mm]{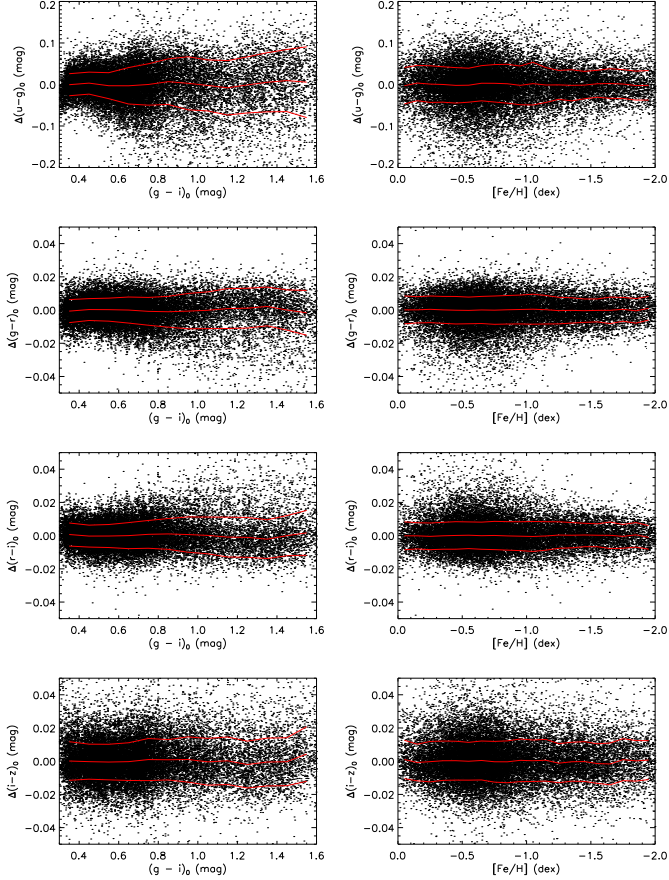}
\caption{
Fit residuals as a function of color $g-i$ and metallicity \feh.
Lines delineating the median and standard deviation of the residuals are over-plotted in red. 
}
\label{}
\end{figure}

\begin{figure}
\includegraphics[width=90mm]{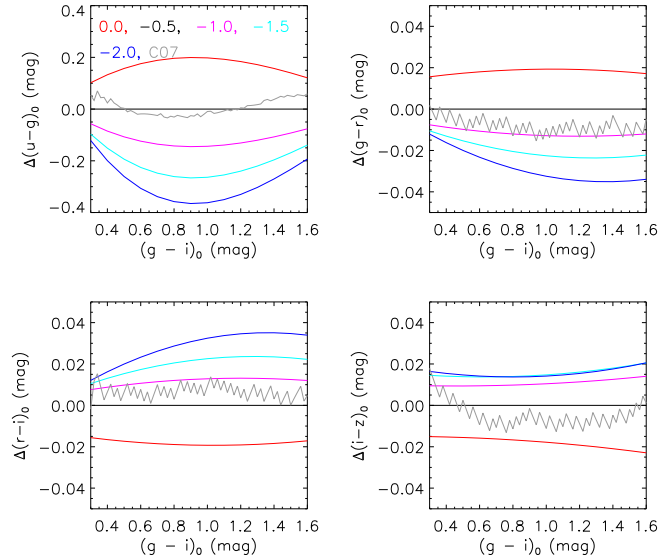}
\caption{
The variations of stellar loci of different metallicities relative to
the corresponding ones of $\feh = -0.5$\,dex.
The stellar loci of Covey et al. (2007) are over-plotted in grey.
}
\label{}
\end{figure}

\begin{table} 
%\begin{minipage}[]{90mm} 
\centering
\caption{Fitting coefficients.}
\label{}
\begin{tabular}{lrrrr} \hline\hline
Coeff. & $u-g^a$ & $g-r^b$ & $r-i^b$ & $i-z^b$  \\
a$_0$ &   1.5003 &   0.0596 &  $-$0.0596 &  $-$0.1060 \\
a$_1$ &   0.0011 &   0.0348 &  $-$0.0348 &  $-$0.0357 \\
a$_2$ &   0.1741 &   0.0239 &  $-$0.0239 &  $-$0.0123 \\
a$_3$ &   0.0910 &   0.0044 &  $-$0.0044 &  $-$0.0017 \\
a$_4$ &   0.0181 &   0.6070 &   0.3930 &   0.2543 \\
a$_5$ &  $-$3.2190 &   0.0261 &  $-$0.0261 &  $-$0.0010 \\
a$_6$ &   1.1675 &  $-$0.0044 &   0.0044 &  $-$0.0050 \\
a$_7$ &   0.0875 &   0.1532 &  $-$0.1532 &  $-$0.0381 \\
a$_8$ &   0.0031 &  $-$0.0136 &   0.0136 &  $-$0.0071 \\
a$_9$ &   7.9725 &  $-$0.0613 &   0.0613 &   0.0030 \\
a$_{10}$ &  $-$0.8190 &  &    &    \\
a$_{11}$ &  $-$0.0439 &  &    &    \\
a$_{12}$ &  $-$5.2923 &  &    &    \\
a$_{13}$ &   0.1339 &    &    &    \\
a$_{14}$ &   1.1459 &    &    &    \\
\hline
\end{tabular}
\begin{description}
\item[$^a$]  $f(x,y)=a_0+a_1*y+a_2*y^2+a_3*y^3+a_4*y^4+a_5*x+$ \\ 
             $ a_6*x*y+a_7*x*y^2+a_8*x*y^3+a_9*x^2+a_{10}*y*x^2+$ \\ 
             $a_{11}*y^2*x^2+a_{12}*x^3+a_{13}*y*x^3+a_{14}*x^4$, where $x$ denotes color $g-i$ and $y$ denotes metallicity \feh. 
\item[$^b$]  $f(x,y)=a_0+a_1*y+a_2*y^2+a_3*y^3+a_4*x+a_5*x*y+a_6*x*y^2+ $ \\ 
             $a_7*x^2+a_8*y*x^2+a_9*x^3$ , where $x$ denotes color $g-i$ and $y$ denotes metallicity \feh.
\end{description}
%\end{minipage} 
\end{table}

The average photometric errors are $0.0073\pm0.0022$
$0.006\pm0.001$, $0.005\pm0.001$, $0.005\pm0.001$, and $0.006\pm0.0015$\,mag in $u, g, r, i$, and $z$ bands, respectively. 
The color calibration uncertainties are 0.005, 0.003, 0.002, and 0.002\,mag for the $u-g$, $g-r$, $r-i$, and $i-z$ colors, 
respectively (Yuan et al. 2014a).
The typical uncertainties of \feh~yielded by the SSPP is about 0.13 dex (Lee et al. 2008b). 
Given the above uncertainties, one expects dispersions of 
0.028, 0.0088, 0.0078, and 0.0085 mag in the $u-g$, $g-r$, $r-i$, and $i-z$ colors, respectively.
The dispersion values yielded by the above fits for the whole selected sample are 
0.029, 0.0076, 0.0076, and 0.0106\,mag, respectively, 
very close or even slightly smaller than the expected values, 
suggesting that the intrinsic widths of the SDSS stellar loci are at maximum a few mmag.
For comparison, the typical widths of loci of Covey et al. (2007) are 0.12, 0.025, 0.025, and 0.035\,mag 
for colors $u-g$, $g-r$, $r-i$, and $i-z$, respectively.
Newberg \& Yanny (1997) find the widths of stellar locus to be smaller than 0.07\,mag 
using The Catalog of $WBVR$ Magnitudes of Northern Sky Bright Stars (Kornilov et al. 1991). 
The large dispersion in the $u-g$ color is mainly contributed by the uncertainties in \feh. 
The dispersions in other colors are dominated by the photometric errors.
In the above analysis, we have neglected the possible effects of variations in the [$\alpha$/Fe] abundance ratio, 
which may contribute parts of the large dispersion in the $u-g$ color.
The possible effects of variable stars have already been largely excluded since only stars flaged as non-variables in 
the Stripe 82 catalog are included in the current analysis. 
Even if some variable stars remain in the current sample, the effects are likely negligible considering that 
the SDSS $u, g, r, i$, and $z$ magnitudes are measured almost simultaneously.
The scatter caused by the uncertainties in reddening corrections, likely smaller than a few mmag given the low 
extinction of the Stripe 82 region, are also ignored.

The measured dispersions are found to be larger for bins of redder $g-i$ color. For example, 
for the bin of color $g-i = 0.4$\,mag and \feh~=$-$1.0\,dex, the measured dispersions are 
0.025, 0.064, 0.064, and 0.01 mag  for colors $u - g$, $g - r$, $r - i$, and $i - z$, respectively. 
While for the bin of color $g-i =0.9$\,mag and \feh~=$-$1.0\,dex, the corresponding values are 
0.052, 0.0094, 0.0094, and 0.012 mag, respectively.
This same phenomenon is seen in Fig.\,2, and it is mainly caused by the fact that stars 
of redder colors are typically fainter and thus tend to have slightly larger photometric errors than stars of bluer colors.
Nevertheless, the measured small dispersions can be fully accounted for by the 
photometric errors, metallicity uncertainties and calibration errors, 
suggesting that the intrinsic widths of the SDSS stellar color loci are at maximum a few mmag. 

We further test the intrinsic widths of the SDSS stellar color loci using open cluster M\,67.
We select a sample of about 400 candidate main sequence stars of M\,67 of colors $0.4 < g-i < 2.0$\,mag based on the photometry of An et al. (2008).
Their $u-g$, $g-r$, $r-i$, and $i-z$ colors are fitted as a function of $g-i$ color. 
The dispersion of the fit residuals are 0.027, 0.008, 0.009, and 0.015\,mag in colors $u-g$, $g-r$, $r-i$, and $i-z$, respectively. 
When fitting the  $u-g$ color, only stars of $u$-band magnitude errors smaller than 0.02\,mag are used.
Given median magnitude errors of 0.009, 0.007, 0.008, 0.008, and 0.010\,mag in the $u, g, r, i$, and $z$ bands, respectively, 
the results support the inference that the intrinsic widths of the SDSS stellar color loci are at maximum a few mmag.
The slightly larger dispersion in $u-g$ color is likely caused by the calibration errors.
For globular clusters, 47\,Tuc for example, although it is known to host 
multiple stellar populations of different chemical compositions (e.g., Anderson et al. 2009; Milone et al. 2012; Li et al. 2014), 
the intrinsic widths of its main sequence stars in the color-magnitude diagram 
is just around 0.01\,mag (see Fig.\,3 of Anderson et al. 2009), which can be reproduced by a dispersion of 0.1 dex in \feh.
The result also supports that the intrinsic widths of the stellar loci of
stars of the same chemical composition must be close to zero.

Thanks to the very small dispersions after accounting for the effects of metallicity,
a prominent feature is revealed: the residuals are non-Gaussian and asymmetric, 
in the sense there is an excess of stars whose colors, compared to what predicted by the fits (cf. Table~1), are bluer in terms of 
$u-g$ and $g-r$ but redder in terms of $r-i$ and $i-z$. 
The asymmetries are most prominent for long wavelength colors (cf. Figs.\,2 and 4).
The asymmetric features can be well explained by binary stars (cf. the companion paper Yuan et al. 2014b).

\begin{figure*}
\includegraphics[width=180mm]{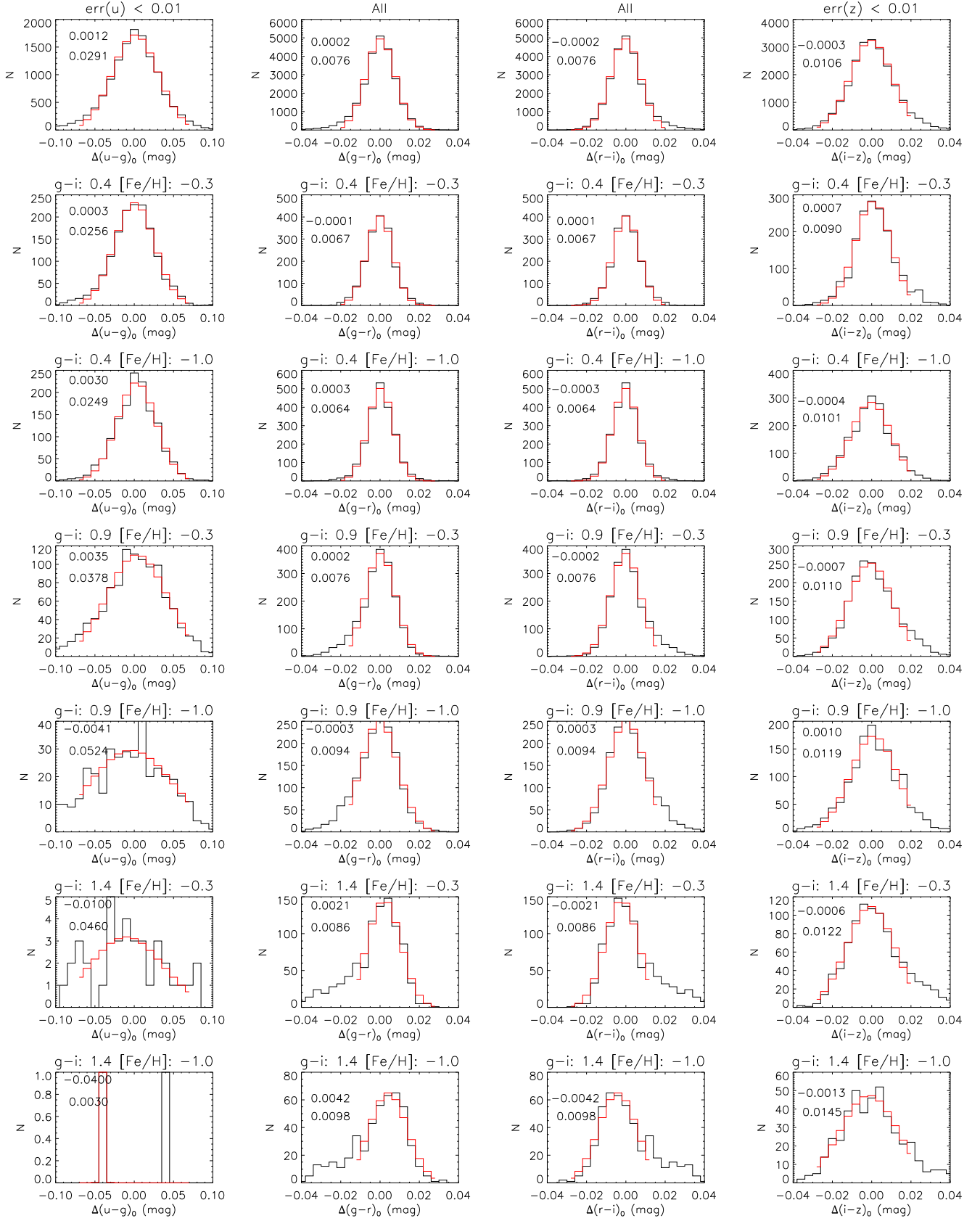}
\caption{
Histogram distributions of fit residuals for 
all sample stars [of $u$ band photometric errors less than 0.01\,mag in the case $(u - g)_0$ and of $i$ band photometric errors less than 0.01\,mag in the case of $(i - z)_0$], and for individual bins of color
$g-i$ and metallicity \feh~bins. 
The values of $g-i$ and \feh~of the bin centers are labelled at the top of each panel. 
The bin widths are 0.4 mag and 0.4 dex in $g - i$ and [Fe/H], respectively. 
The Gaussian fits to the distributions are over-plotted in red. 
The offset and dispersion of the fit are labelled. Note the residual distributions are non-Gaussian and asymmetric. 
In the bottom left panel, the number of stars is too small to yield a reasonable fit. 
}
\label{}
\end{figure*}

\section{Conclusions and discussions}
In this paper,  by combining the spectroscopic information and re-calibrated imaging photometry of 
the SDSS Stripe 82, we have built a large, clean sample of main sequence stars with accurate colors 
and  well determined metallicities
to study the metallicity dependence and intrinsic widths of the SDSS stellar loci.
The colors vary typically by 0.20, 0.02, $-$0.02, and $-$0.02 mag 
per metallicity dex for the $u-g$, $g-r$, $r-i$, and $i-z$ colors, respectively.
The variations are larger for metal-rich stars than for metal-poor ones, 
and for F/G/K stars than for A/M ones.
Using the sample, we have performed two dimensional polynomial fitting to the $u-g$, $g-r$, $r-i$, and $i-z$ colors 
as a function of the color $g-i$ and metallicity \feh.
We find that the $u-g$, $g-r$, $r-i$, and  $i-z$ colors of main sequence stars are 
accurately determined by their $g-i$ colors and metallicities. 
The fit residuals, at the level of 0.029, 0.008, 0.008, and 0.011\,mag for the $u-g$, $g-r$, $r-i$, and $i-z$ colors respectively,
are fully accounted for by the photometric errors, metallicity uncertainties and calibration errors, suggesting that the intrinsic widths
of the loci are at maximum a few mmag.

By taking into the metallicity dependence of stellar loci into account,
the scattering of the loci is much reduced. As a direct consequence, this reveal remarkable  
asymmetric distributions of residuals of fits to the stellar loci, 
which in turn suggests that a significant fraction of stars are in binaries.
The asymmetric residual distributions provide a unique opportunity to obtain an unbiased estimate of the 
binary fraction for field stars. 
The results will be presented in a companion paper (Yuan et al. 2014b). 

The metallicity dependent stellar loci provide a much more precise definition of the distribution of ``normal" stars in the 
color-color space, and consequently a much more precise way to select exotic objects as  ``outliers", 
such as quasars, IR/UV excess objects, giant stars, white dwarfs, and white-dwarf-main-sequence binaries.
They will also discriminate galaxies and globular clusters from stars in deep multi-band imaging surveys where 
the galaxies and globular clusters can not be spatially resolved. 

The metallicity dependent stellar loci will also benefit accurate determinations of  
stellar parameters, such as photometric metallicities for stars of a wide spectral type from the late-A to the early-M.
Compared to the photometric-metallicity relation of Ivezi{\'c} et al. (2008),
the metallicity dependent stellar loci are capable of delivering photometric metallicities of remarkable precision 
about 0.13\,dex for the sample stars of the current work, depending on metallicities and colors of the stars, 
and can be applied to stars of a wider range of colors and metallicities.
In the third paper of this series (Yuan et al. 2014c, in preparation), accurate estimates of 
photometric metallicities will be presented for about 0.5 million stars in Stripe\,82. 
When combined with IR photometry data such as the 2MASS (Skrutskie et al. 2006) or the WISE (Wright et al. 2010) surveys,
one is able to determine values of the interstellar reddening and metallicity for individual stars 
using the metallicity dependent stellar loci,
opening up a possibility of topographical studies of the Galactic disk(s) (Yuan et al. 2014, in prep)
using the huge data sets of deep photometric surveys already available or forthcoming.
The metallicity dependent stellar loci also serve as a natural ingredient for the Bayesian approach of  
stellar parameter determinations that makes use of the constraints from the multi-band photometry 
as well as spectroscopy (e.g., Schoenrich \& Bergemann 2014; Green et al. 2014; Zhang et al. 2014, in prep).

The above potential applications of metallicity dependent stellar loci rely on the accurate color calibration. 
Future attempts of calibration based on 
the stellar locus regression method (SLR; High et al. 2009) for color calibration should fully take the metallicity 
effects into account to minimize possible systematic calibration errors. 
By properly and fully taking into account of the metallicity effects, 
the stellar color regression (SCR) method, proposed by Yuan et al. (2014a), for example,
is able to achieve a color calibration accuracy approaching a few mmag. 

\vspace{7mm} \noindent {\bf Acknowledgments}{
We would like to thank the referee for his/her valuable comments.
This work is supported by National Key Basic Research Program of China 2014CB845700 
and China Postdoctoral Science special Foundation 2014T70011.
This work has made use of data products from the Sloan Digital Sky Survey.
}

\end{document}